\documentclass{jetpl}
\twocolumn

\lat
\usepackage[dvips]{graphicx}
\usepackage{epsfig}
\usepackage{amsmath}
\usepackage{amssymb}
\usepackage{amsfonts}

\newcommand{\IR}{{\mathrm{IR}}}
\newcommand{\UV}{{\mathrm{UV}}}

\newcommand{\bea}{\begin{eqnarray}}
\newcommand{\ena}{\end{eqnarray}}
\newcommand{\dd}{{\mathrm d}}

\title{Correlation functions of descendants in the scaling \\
Lee--Yang model}

\rtitle{Correlation functions of descendants in the scaling Lee--Yang model}

\author{V.\,A.\,Belavin$^+$\/\thanks{belavin@itep.ru},
O.\,V.\,Miroshnichenko$^{*}$ \\
$^{+}$ITEP, B.~Cheremushkinskaya 25, Moscow, 117259, Russia\\
$^{*}$L.D.Landau Institute for Theoretical Physics,  Chernogolovka, 142432,
Russia}

\rauthor{V.\,A.\,Belavin, O.\,V.\,Miroshnichenko}

\date{}

\abstract{Correlation functions of the composite field $T\bar{T}$ in
the scaling Lee--Yang model are studied. Using the analytic expression for
form factors of this operator recently proposed by
Delfino and Niccoli~\cite{DN}, we show numerically that the constraints on the $T\bar{T}$
expectation values obtained in~\cite{AZ_VEVTT} and the additional requirement
of asymptotic behavior lead to a perfect agreement with the ultraviolet
asymptotic predicted by the conformal perturbation theory.}

\PACS{11.25.Hf}

\begin{document}

\maketitle

In the present work, we use two different approaches to massive integrable 
quantum field theories. 
In the first approach, the massive theory is considered as a
perturbation of a certain conformal field theory (being a fixed
point of the renormalization group flow) by a relevant
operator~\cite{Z-Int}. The structure of the space
of local operators does not change along the renormalization group
flow; therefore, the space of local operators of the massive theory
is assumed to be isomorphic to the corresponding conformal field theory space.
This space consists of primary operators and descendants~\cite{BPZ}. 
The correlation functions are calculated using operator product
expansions
\begin{equation}
\left\langle A_{i}\left( x\right) A_{j}\left( 0\right) \right\rangle
=\sum_{k}C_{ij}^{k}\left( x,g\right) \left\langle A_{k}\right\rangle \,,
\label{2}
\end{equation}
where $g$ is the perturbing coupling constant.
The structure functions of the operator
algebra $C_{ij}^{k}\left( x,g\right) $, are assumed to be analytic in $g$, 
provided that the renormalized 
fields $A_{i}$ are chosen to have definite dimensions.
This hypothesis provides the way to expand the structure functions in a 
perturbation series in the coupling constant. In the zeroth order,
the $C_{ij}^{k}\left( x,g\right) $ coincide with the
structure functions of conformal field theory. This
procedure, called Conformal Perturbation
Theory, was invented in~\cite{Z-LY}. The vacuum expectation values $%
\left\langle A_{i}\right\rangle $ depend on $g$ nonanalytically
and cannot be calculated using perturbation theory. It follows
from dimensional
analysis that%
\begin{equation*}
\left\langle A_{i}\right\rangle =Q_{i}g^{\frac{\Delta _{i}}{1-\Delta }} \,,
\end{equation*}
where $Q_{i}$ is independent of $g$, $\Delta_{i}$ is the dimension of the field
$A_{i}$, and $\Delta$ is the dimension of the perturbation.
The vacuum expectation values of primary
fields and the first nontrivial descendants were obtained in~\cite{LZ,FFLZZ,FLZZ}. 
Taking into account that $\Delta<1$ and the dimensions $\Delta_{i}$ are increasing,
expression~(\ref{2}) represents the series in increasing powers of $x$,
and we can restrict ourself to a small number of terms for $x<<1$.

In the second (form-factor) approach, 
matrix elements of local
operators in the basis of asymptotic states are
determined from the Smirnov axioms~\cite{Smirnov}, given the $S$-matrix
and mass spectrum.
Correlation functions then can be represented as a spectral decomposition.

We consider
the scaling Lee--Yang model
\begin{equation}
S=S_{M\left( 2/5\right) }+g\int \varphi \,\dd^{2}x \,,  \label{3}
\end{equation}%
i.e., the $M\left( 2/5\right) $ minimal model of
conformal field theory perturbed by the field $\varphi =\phi _{1,3}$,
which is the only nontrivial primary field in it. The model
$M\left( 2/5\right) $ has the central charge $c=-22/5$.
The space of fields in this model consists of two
primary fields: the identity operator $I=\phi
_{1,1}=\phi _{1,4}$ and the field $\varphi =\phi _{1,2}=\phi
_{1,3}$ with the right and left dimensions $\Delta
=\bar{\Delta}=-1/5$ and their descendants. 
It is convenient to
consider the trace of the stress tensor $\Theta $,
\begin{equation}
\Theta \left( x\right) =\pi g\left( 1-\Delta \right) \varphi \left(
x\right) \,.  \label{8}
\end{equation}%
The vacuum expectation value of $\left\langle \Theta \right\rangle $ was
obtained in~\cite{Z-TBA}
\begin{equation}
\left\langle \Theta \right\rangle 
=-\frac{\pi m^{2}}{4\sqrt{3}}  \,,\label{9}
\end{equation}%
where $m$ is the particle mass. In what follows, we also use the relation
\begin{equation}
g=i\frac{2^{\frac{1}{5}}5^{\frac{3}{4}}}{16\pi ^{\frac{6}{5}}}\frac{\left(
\Gamma \left( \frac{2}{3}\right) \Gamma \left( \frac{5}{6}\right) \right) ^{%
\frac{12}{5}}}{\Gamma \left( \frac{3}{5}\right) \Gamma \left( \frac{4}{5}%
\right) }m^{\frac{12}{5}}
\label{g-m}
\end{equation}%
between the coupling $g$ and the  scale $m$ of the theory,
found in~\cite{Z-TBA,Z-RSG}.

The form factors of the operator $\left\langle \Theta \right\rangle$
were found in~\cite{Z-LY}. The form factors of the operator
$T\overline{T}=L_{-2}\overline{L}_{-2}I$ in this model were
recently obtained in~\cite{DN}. We use this expression to
numerically calculate the correlation functions 
$G\left( m\left\vert x\right\vert \right) =m^{-6} \left\langle T\overline{T}%
\left( x\right) \Theta \left( 0\right) \right\rangle $ and 
$H\left( m\left\vert x\right\vert \right) =m^{-8}\left\langle T%
\overline{T}\left( x\right) T\overline{T}\left( 0\right) \right\rangle $ up
to three-particle terms in spectral expansions, and we compare them with the
leading terms in conformal perturbation theory~(\ref{2}).

The first three nonzero vacuum expectation values in expansion~(\ref{2}) 
are $\left\langle \varphi \right\rangle $, $\left\langle I\right\rangle =1$, 
and $\left\langle L_{-2}\overline{L}_{-2}I\right\rangle 
=\left\langle T\overline{T}\right\rangle $. 
We recall the operator product expansions of the stress tensor
in a conformal field theory~\cite{BPZ}:
\bea
T\left( z\right) \phi \left( 0\right) &=& \frac{\Delta }{z^{2}}\phi \left(
0\right) +\frac{1}{z}\partial \phi \left( 0\right) +\ldots  \,, \label{4} \\
T\left( z\right) T\left( 0\right) &=&\frac{c}{2z^{4}}+\frac{2}{z^{2}}T\left(
0\right) +\frac{1}{z}\partial T\left( 0\right) +\ldots \,.  \label{5}
\ena
This gives 
\begin{equation*}
C_{T\overline{T,}\varphi }^{I}\left( x\right) =0,\quad C_{T\overline{T,}%
\varphi }^{\varphi }\left( x\right) =\Delta ^{2}\left\vert x\right\vert
^{-4},
\end{equation*}%
\begin{equation*}
\quad C_{T\bar{T},T\bar{T}}^{I}\left( x\right) =\left( \frac{c}{2}\right)
^{2}\left\vert x\right\vert ^{-8},\quad C_{T\bar{T},T\bar{T}}^{\varphi
}\left( x\right) =0 \,.
\end{equation*}
We here skip the details of the first-order calculations for the structure
functions. All nessesary formulas can be found
in~\cite{Z-LY,BBLPZ}.
 Using~(\ref{8}) and~(\ref{9})  
we finally obtain the expression
for $\left\langle T\overline{T}\left( x\right) \Theta\left( 0\right)
\right\rangle $:
\begin{equation*}
G_{\UV}\left( m\left\vert x\right\vert \right)=
-\frac{\pi }{100\sqrt{3}}\left( m\left\vert x\right\vert \right)
^{-4}+g_{1}\left(2\ln \mu \left\vert x\right\vert + \right.
\end{equation*}
\begin{equation}
\left.+g_{2}\right)\left( m\left\vert x\right\vert \right) ^{-\frac{8}{5}} +g_{3}\left( 2\ln \mu \left\vert x\right\vert +
g_{4}\right)\left( m\left\vert x\right\vert \right)^{-\frac{6}{5}} \,,
\label{10}
\end{equation}%
\bea
g_{1}&=&-\frac{3^{\frac{3}{2}}\pi ^{\frac{4}{5}}\left( \Gamma \left( \frac{2}{3%
}\right) \Gamma \left( \frac{5}{6}\right) \right) ^{\frac{12}{5}}\left(
\Gamma \left( \frac{1}{5}\right) \right) ^{\frac{3}{2}}\left( \Gamma \left( 
\frac{2}{5}\right) \right) ^{\frac{1}{2}}}{2^{\frac{19}{5}}5^{\frac{17}{4}%
}\left( \Gamma \left( \frac{3}{5}\right) \right) ^{\frac{3}{2}}\left( \Gamma
\left( \frac{4}{5}\right) \right) ^{\frac{5}{2}}} \,, \nonumber \\
g_{2}&=&2\psi \left( 2\right) -\psi \left( -\frac{1}{5}\right) -\psi \left( 
\frac{1}{5}\right) -\frac{115}{18} \,, \nonumber \\
g_{3}&=&\frac{27}{\pi ^{\frac{2}{5}}2^{\frac{13}{5}}5^{\frac{7}{2}}}\frac{%
\left( \Gamma \left( \frac{2}{3}\right) \Gamma \left( \frac{5}{6}\right)
\right) ^{\frac{24}{5}}}{\left( \Gamma \left( \frac{3}{5}\right) \Gamma
\left( \frac{4}{5}\right) \right) ^{2}} \,, \nonumber \\
g_{4}&=&2\psi \left( 2\right) -\psi \left( -\frac{2}{5}\right) -\psi \left( 
\frac{2}{5}\right) -\frac{55}{14} \,. \nonumber
\ena

In the same way 
for $\left\langle T\overline{T}\left( x\right) 
T\overline{T}\left( 0\right) \right\rangle$, we have
\begin{equation*}
H_{\UV} \left( m\left\vert x\right\vert \right) =
\left( \frac{c}{2}\right) ^{2}\left(m\left\vert x\right\vert\right) ^{-8}
+\frac{\pi \Delta ^{2}\left( 1-\Delta \right) }{\sqrt{3}}
\times
\end{equation*}
\begin{equation}
\times\left( 4\ln \mu \left\vert x\right\vert +\frac{c-4\Delta
^{2}}{2\Delta \left( \Delta -1\right) }\right) 
\left(m\left\vert x\right\vert\right)^{-6} \,.
\label{6}
\end{equation}
In the first-order calculations, we have faced the resonance
problem~\cite{Z-LY}, leading to the undefined coefficient $\mu$
in the subleading terms in~(\ref{10}) and~(\ref{6}).

The mass spectrum of the scaling Lee--Yang model consists of one
particle $A$. Correlation functions can be
expressed through the form factors of local operators as spectral sums.
For example, the two-point Euclidean correlation function of the operators
$\mathcal{O}_{1}$ and $\mathcal{O}_{2}$ has the form
\begin{equation*}
\left\langle \mathcal{O}_{1}\left( x\right) \mathcal{O}_{2}\left( 0\right)
\right\rangle =
\end{equation*}%
\begin{equation*}
=\sum_{n=0}^{\infty }\int \frac{\dd\theta _{1}\ldots \dd\theta _{n}}{n!\left(
2\pi \right) ^{n}}F_{n}^{\mathcal{O}_{1}}\left( \theta _{1},\ldots \theta
_{n}\right) \times
\end{equation*}%
\begin{equation}
\times F_{n}^{\mathcal{O}_{2}}\left( \theta _{1}-i\pi ,\ldots \theta
_{n}-i\pi \right) e^{-m\left\vert x\right\vert \sum\limits_{i=1}^{n}\cosh
\theta _{i}}  \,. \label{20}
\end{equation}%
The expressions for the first four form factors of the operator
$\Theta $ have the
forms~\cite{Z-LY}:%
\bea
F_{0}^{\Theta }&=&-\frac{\pi m^{2}}{4\sqrt{3}} \,, \label{30} \\
F_{1}^{\Theta }&=&-\frac{i\pi m^{2}}{2^{\frac{5}{2}}3^{\frac{1}{4}}v\left(
0\right) }  \,, \label{31} \\
F_{2}^{\Theta }\left( \theta _{1},\theta _{2}\right) &=& \frac{\pi m^{2}}{2}%
\frac{f\left( \theta _{1}-\theta _{2}\right) }{4v^{2}\left( 0\right) } \,,
\label{32} \\
F_{3}^{\Theta }\left( \theta _{1},\theta _{2},\theta _{3}\right) &=& \nonumber
\ena
\begin{equation}
=\frac{i3^{\frac{1}{4}}\pi m^{2}}{2^{\frac{7}{2}}v^{3}\left( 0\right) }%
\prod\limits_{i<j}^{3}f\left( \theta _{i}-\theta _{j}\right) \left( 1+\frac{1%
}{8\prod\limits_{i<j}\cosh\frac{\theta _{i}-\theta _{j}}{2}}\right) \,,
\label{33}
\end{equation}%
where
\begin{equation}
f\left( \theta \right) =\frac{\cosh\theta -1}{\cosh\theta +1/2}%
v\left( i\pi -\theta \right) v\left( -i\pi +\theta \right)  \,, \label{22}
\end{equation}%
\begin{equation}
v\left( \theta \right) =\exp \left( 2\int\limits_{0}^{\infty }\frac{\sinh%
\frac{t}{2}\sinh\frac{t}{3}\sinh\frac{t}{6}}{t\sinh^{2}t}e^{%
\frac{i\theta t}{\pi }}\dd t\right)  \,. \label{23}
\end{equation}%
The following expression for the form factors of the operator
$T\overline{T}$ was obtained in~\cite{DN} using the restriction on the
growth at infinity, the asymptotic factorization properties, and
the relation for the expectation value of $T\overline{T}$ obtained 
in~\cite{AZ_VEVTT}:
\begin{equation*}
F_{n}^{T\overline{T}}=m^{2}\left( a\left( \sigma _{1}^{\left( n\right) }\bar{%
\sigma}_{1}^{\left( n\right) }\right) ^{2}+c\sigma _{1}^{\left( n\right) }%
\bar{\sigma}_{1}^{\left( n\right) }+d\right) F_{n}^{\Theta }+
\end{equation*}%
\begin{equation}
+bF_{n}^{\mathcal{K}_{3}}+em^{4}\delta _{n,0}  \,, \label{34}
\end{equation}%
where%
\begin{equation}
\sigma _{1}^{\left( n\right) }=\sum_{i=1}^{n}x_{i},\quad \bar{\sigma}%
_{1}^{\left( n\right) }=\sum_{i=1}^{n}\frac{1}{x_{i}} \,. \label{35}
\end{equation}%
For $n<3$, $F_{n}^{\mathcal{K}_{3}}$ is the solution equal to zero,
and at $n=3$,%
\begin{equation*}
F_{n}^{\mathcal{K}_{3}}\left( \theta _{1},\theta _{2},\theta _{3}\right) 
=-i\left( \frac{3}{4}\right) ^{\frac{3}{4}}\frac{m^{2}}{v^{3}\left( 0\right)}
\times
\end{equation*}%
\begin{equation}
\times\prod\limits_{i<j}^{3}f\left( \theta _{i}-\theta _{j}\right) \left( \cosh
\left( \theta _{i}-\theta _{j}\right) +\frac{1}{2}\right)  \,. \label{36}
\end{equation}%
The constants $a$, $b$, $d$, are $e$ are
\begin{equation}
a=\frac{\left\langle \Theta \right\rangle }{m^{2}},\quad b=-\frac{%
\left\langle \Theta \right\rangle ^{2}}{m^{4}},  \label{37-1}
\end{equation}%
\begin{equation}
\quad d=-\frac{2}{m^{2}}\left\langle \Theta \right\rangle ,\quad e=-\frac{%
\left\langle \Theta \right\rangle ^{2}}{m^{4}} \,, \label{37-2}
\end{equation}
where
\begin{equation}
\left\langle \Theta \right\rangle =-\frac{\pi m^{2}}{4\sqrt{3}} \,. \label{38}
\end{equation}%
The constant $c$ is not determined, which corresponds to an ambiguity $T%
\overline{T}\rightarrow T\overline{T}+ \#\partial \overline{\partial }%
\varphi $ in the definition of the operator $T\overline{T}$ outside
the critical point, because the dimensions of operators $T\overline{T}$ and
$\partial \overline{\partial }\varphi $ satisfy the resonance
condition~\cite{Z-LY}. The coefficient $a$ is determined only from the
restriction on the growth at infinity and the asymptotic factorization
condition, and the coefficients $d$ and $e$, from the growth restriction
and Zamolodchikov relation for the stress-tensor expectation value. The
coefficient $b$ is determined from each of these sets of conditions
independently.

Formulas~(\ref{30})--(\ref{33}),~(\ref{34}), and~(\ref{36}) lead
to the spectral expansion of the correlation functions
$G\left( m\left\vert x\right\vert \right)$ and
$H\left( m\left\vert
x\right\vert \right)$ up to three-particle terms:%
\begin{equation*}
G_{\IR}\left( m\left\vert x\right\vert \right) =
\left( \frac{\pi }{4\sqrt{3}}\right) ^{3} +
\end{equation*}%
\begin{equation}
+G_{1}\left( m\left\vert
x\right\vert \right) +G_{2}\left( m\left\vert x\right\vert \right)
+G_{3}\left( m\left\vert x\right\vert \right) +\ldots   \,\,, \label{39}
\end{equation}%
\begin{equation*}
H_{\IR}\left( m\left\vert x\right\vert \right) 
=\left( \frac{\pi }{4\sqrt{3}}\right) ^{4} +
\end{equation*}%
\begin{equation}
+H_{1}\left( m\left\vert
x\right\vert \right) +H_{2}\left( m\left\vert x\right\vert \right)
+H_{3}\left( m\left\vert x\right\vert \right) +\ldots \,\,,  \label{40}
\end{equation}%
where
\bea
G_{1}\left( x\right) &=& -\frac{\pi }{32\sqrt{3}v^{2}\left( 0\right) }
\left(a+c+d\right) K_{0}\left( x\right) \,,\label{42}\\
G_{2}\left( x\right) &=&
\frac{1}{128v^{4}\left( 0\right) }\int\limits_{0}^{\infty } \bigg( 
4a\left(1+\cosh\theta \right)^{2}+2c\left( 1+ \right. 
\nonumber \\ 
&& \left. \cosh\theta \right)+ d \,\bigg)  
g\left( \theta \right) K_{0}\left( 2x\cosh\frac{\theta }{2}\right)
\dd\theta  \,, \label{43} \\ 
G_{3}\left( x\right) &=&
\frac{1}{32\pi ^{2}v^{6}\left( 0\right) }\int\limits_{0}^{\infty
}\int\limits_{0}^{\infty }\left( -\frac{\pi }{32\sqrt{3}}B\left( \theta,
\chi \right) \times \right. \nonumber 
\ena 
\begin{equation*}
\left( a\,A^{4}\left( \theta ,\chi \right)
 +  c\,A^{2}\left( \theta ,\chi \right) +d\right) 
 +b\,C\left( \theta,\chi \right) \bigg) B \left( \theta ,\chi \right)\times  \nonumber
\end{equation*}
\begin{equation}
\times  g\left( \theta \right) g\left( \chi
\right) g\left( \theta -\chi \right) K_{0}\left( A\left( \theta ,\chi
\right) x\right) \dd\theta \dd\chi \,,
\label{44}
\end{equation}
\bea
H_{1}\left( x\right) &=& -\frac{\pi }{32\sqrt{3}v^{2}\left( 0\right) }\left(
a+c+d\right) ^{2}K_{0}\left( x\right) \,, \label{45} \\
H_{2}\left( x\right) &=&
\frac{1}{128v^{4}\left( 0\right) }\int\limits_{0}^{\infty }\bigg( 4a\left(
1+\cosh\theta \right) ^{2} +2c\left( 1+ \right.  \nonumber \\
&&\left. +\cosh\theta \right) +d\bigg)^2 g\left( \theta \right) K_{0}\left( 2x\cosh\frac{\theta }{2}
\right) \dd\theta   \,, \label{46} \\
H_{3}\left( x\right) &=&
-\frac{\sqrt{3}}{\pi ^{3}v^{6}\left( 0\right) }\int\limits_{0}^{\infty
}\int\limits_{0}^{\infty }\bigg( -\frac{\pi }{32\sqrt{3}}B\left( \theta
,\chi \right) \times \nonumber 
\ena
\begin{equation*}
\times \left( aA^{4}\left( \theta ,\chi \right) 
 + cA^{2}\left( \theta ,\chi \right) +d\right) +
b\,C\left( \theta,\chi \right) \bigg) ^{2}\times 
\end{equation*}
\begin{equation}
\times g\left( \theta \right) g\left( \chi \right) g\left( \theta -\chi
\right) K_{0}\left( A\left( \theta ,\chi \right) x\right) 
\dd\theta \dd\chi \,,
\bigskip   \label{47}
\end{equation}
\begin{equation}
g\left( \theta \right) =f\left( \theta \right) f\left( -\theta \right) \,,
\label{48}
\end{equation}%
\bea
A\left( \theta ,\chi \right) &=& \sqrt{3+2\left( \cosh\theta +\cosh\chi
+\cosh\left( \theta -\chi \right) \right) } \,, \nonumber \\
B\left( \theta ,\chi \right) &=& 1+\frac{1}{8\cosh\frac{\theta }{2}\cosh
\frac{\chi }{2}\cosh\frac{\theta -\chi }{2}} \,, \nonumber \\
C\left( \theta ,\chi \right) &=& 
\left( \cosh\theta +\frac{1}{2}\right) \left( \cosh\chi +\frac{1}{2}
\right) 
\times \nonumber \\ &&
\times \left( \cosh\left( \theta -\chi \right) +\frac{1}{2}\right) \,.
\nonumber
\ena

\begin{figure}[!htb]
\begin{center}
\begin{tabular}{cl}
\includegraphics[angle=-00,width=8.5cm,height=6.5cm,clip=true]{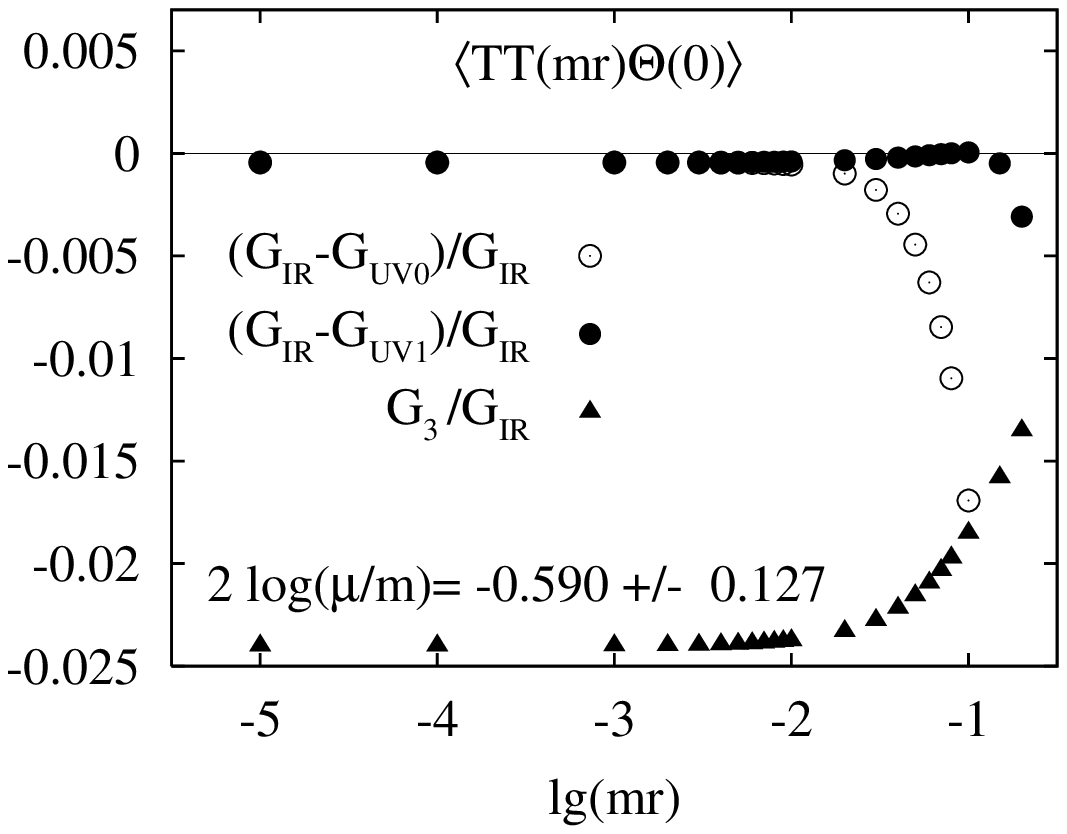} &
\\ 
\includegraphics[angle=-00,width=8.5cm,height=6.5cm,clip=true]{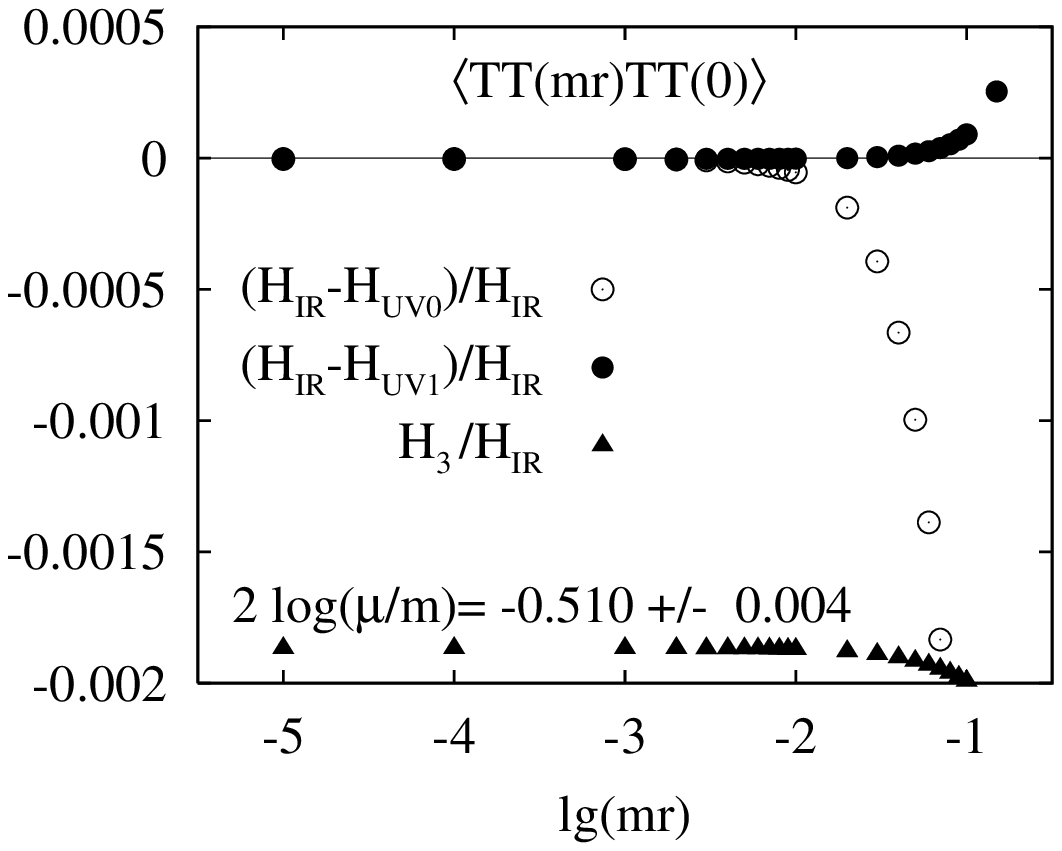} &
\end{tabular}
\end{center}
\caption{Fig.~1: The results of fitting the IR data for $c=0$
by the UV expansions with the fitting parameter $\mu$.}
\label{fig:examples}
\end{figure}

As mentioned above, the ambiguity in the definition of the 
operator $T\overline{T}$ does not affect the leading UV order.
Nevertheless, it is interesting to establish the exact correspondence between the coefficients
$\mu$ and $c$. It happens that the precision of the form-factor calculations is
insufficient for this purpose.

In Fig.1, we show the results of fitting
IR data obtained for $c=0$ by the UV expansions.
In view of the rapid increase of the correlation functions considered,
we use a logarithmic scale and also the ratios of the corresponding
contributions for better visibility.
The ratios of the three-particle contributions $G_3/G_{\IR}$ and
$H_3/H_{\IR}$ (triangles) are given in the figure to visibly demonstrate
the degree of agreement between the IR and UV data and also to estimate
in which interval we expect the IR data to be valid.
As a result of such an estimation, we fitted on the interval $[0.01,0.2]$.
Further, we show the degree of agreement of the IR data
with the zeroth-order (open circles) and
with the first-order (filled circles) UV expansions.
The errors in determining the fitting parameters shows that
we must take the higher-order
form-factor contributions into account.

A comparison of the numerical values 
of UV expansions~(\ref{10}) and~(\ref{6}) with
form-factor expansions~(\ref{39}) and~(\ref{40})
up to three-particle contributions for
$10^{-5}<m\left\vert x\right\vert <0.2$ is shown in Tables~$1$ and~$2$.
In the IR expansion, we use $c=0$.
The results of the first-order UV calculations are given for the best-fit
value of $2\log(\mu / m)=-0.51$.
It can be seen that the UV and IR expansions
coincide with sufficiently good accuracy.

In conclusion, we emphasize that this comparison
confirms the construction for the form factors of operator 
$T\overline{T}$ proposed in~\cite{DN}.
Indeed, we note that the two-particle terms $G_{2}$ and $H_{2}$ are
highly sensitive to the value of the parameter $a$, and the
three-particle terms $G_{3}$ and $H_{3}$, to the value of the
parameter $b$ (because the corresponding terms in the integrands
in~(\ref{43}),~(\ref{44}),~(\ref{46}), and~(\ref{47}) have the greatest
increase at infinity) and are weakly sensitive to the values of the other
parameters. Therefore,
the fact that the sums of the first three terms of the form-factor
expansions (i.e., zero-, one-, and two-particle) coincide with
the UV expansion with an accuracy up to three digits
confirms the value of the parameter $a$ and the assumption of
asymptotic behavior for descendent operators in the scaling
Lee--Yang model~\cite{DN}.
Including three-particle terms improves the convergence up to five
digits
which confirms the value of the parameter $b$ and all conjectures
used to determine it~\cite{DN,AZ_VEVTT}.

The authors are grateful to A.~A.~Belavin for the useful discussions. This
work was supported by the following grants: RFBR 04-02-16027, SS
2044.2033.2, INTAS-03-51-3350. V.~B.~acknowledges the hospitality and
stimulating scientific atmosphere at LPTHE in Jussieu.


\onecolumn

\begin{table}[tbp] \centering%

$%
\begin{tabular}{|c|c|c|c|c|}
\hline
$m\left\vert x\right\vert $ & $G_{\IR}$ up to 2 particles & $G_{\IR}$ up to 3
particles & ~~$G_{\UV}$ leading term~~ & ~~~$G_{\UV}$ first order~~~ \\
\hline
0.00001&-1.85653286e+18&-1.81300206e+18&-1.81380007e+18&-1.81380007e+18\\ 
0.00010&-1.85653275e+14&-1.81300214e+14&-1.81380007e+14&-1.81380007e+14\\ 
0.00100&-1.85652408e+10&-1.81300401e+10&-1.81380007e+10&-1.81379845e+10\\ 
0.00200&-1.16031262e+09&-1.13312746e+09&-1.13362505e+09&-1.13362023e+09\\ 
0.00400&-7.25161988e+07&-7.08199961e+07&-7.08515654e+07&-7.08501625e+07\\ 
0.00600&-1.43231670e+07&-1.39888795e+07&-1.39953709e+07&-1.39946955e+07\\ 
0.00800&-4.53151109e+06&-4.42603060e+06&-4.42822284e+06&-4.42782279e+06\\ 
0.01000&-1.85589100e+06&-1.81282105e+06&-1.81380007e+06&-1.81353444e+06\\ 
0.02000&-1.15890841e+05&-1.13251558e+05&-1.13362505e+05&-1.13289636e+05\\ 
0.03000&-2.28615882e+04&-2.23527759e+04&-2.23925935e+04&-2.23590156e+04\\ 
0.04000&-7.22104580e+03&-7.06438900e+03&-7.08515654e+03&-7.06596919e+03\\ 
0.05000&-2.95154366e+03&-2.88923575e+03&-2.90208012e+03&-2.88972936e+03\\ 
0.06000&-1.41992160e+03&-1.39078944e+03&-1.39953709e+03&-1.39095957e+03\\ 
0.07000&-7.64321784e+02&-7.49094702e+02&-7.55435266e+02&-7.49155160e+02\\ 
0.08000&-4.46655159e+02&-4.38020065e+02&-4.42822284e+02&-4.38041748e+02\\ 
0.10000&-1.81664934e+02&-1.78361775e+02&-1.81380007e+02&-1.78368395e+02\\ 
0.15000&-3.50788352e+01&-3.45331307e+01&-3.58281496e+01&-3.45550348e+01\\ 
0.20000&-1.07663389e+01&-1.06227468e+01&-1.13362505e+01&-1.06572736e+01\\ 
\hline
\end{tabular}%
$

\caption{Table 1: Numerical data for the correlation function
$\left\langle T\bar{T}\left( x\right) \Theta \left( 0\right)\right\rangle $.}
\label{Tab1}%
\end{table}%

\begin{table}[tbp] \centering%

\begin{tabular}{|c|c|c|c|c|}
\hline
$m\left\vert x\right\vert $ & $H_{\IR}$ up to 2 particles & $H_{\IR}$ up to 3
particles & ~~$H_{\UV}$ leading term~~ & ~~~$H_{\UV}$ first order~~~ \\
\hline
0.00001&4.84902564e+40&4.83998645e+40&4.84000118e+40&4.84000118e+40\\ 
0.00010&4.84902561e+32&4.83998641e+32&4.84000118e+32&4.84000114e+32\\ 
0.00100&4.84901983e+24&4.83998034e+24&4.83999827e+24&4.83999495e+24\\ 
0.00200&1.89414518e+22&1.89061384e+22&1.89062440e+22&1.89061959e+22\\ 
0.00400&7.39895858e+19&7.38516043e+19&7.38525142e+19&7.38518211e+19\\ 
0.00600&2.88693111e+18&2.88154509e+18&2.88161097e+18&2.88155315e+18\\ 
0.00800&2.89015215e+17&2.88475727e+17&2.88486384e+17&2.88476475e+17\\ 
0.01000&4.84879446e+16&4.83973781e+16&4.84000106e+16&4.83974909e+16\\ 
0.02000&1.89382175e+14&1.89026924e+14&1.89062538e+14&1.89026936e+14\\ 
0.03000&7.38795913e+12&7.37402153e+12&7.37692595e+12&7.37399344e+12\\ 
0.04000&7.39439146e+11&7.38034934e+11&7.38525556e+11&7.38027952e+11\\ 
0.05000&1.24017793e+11&1.23780576e+11&1.23903991e+11&1.23778474e+11\\ 
0.06000&2.88317563e+10&2.87761876e+10&2.88161166e+10&2.87754224e+10\\ 
0.07000&8.39672143e+09&8.38041127e+09&8.39577991e+09&8.38008907e+09\\ 
0.08000&2.88379417e+09&2.87814821e+09&2.88486545e+09&2.87799616e+09\\ 
0.09000&1.12334075e+09&1.12112407e+09&1.12435943e+09&1.12104556e+09\\ 
0.10000&4.83278636e+08&4.82317554e+08&4.83999982e+08&4.82273987e+08\\ 
0.15000&1.87883587e+07&1.87496455e+07&1.88849209e+07&1.87448871e+07\\ 
0.20000&1.87207209e+06&1.86810743e+06&1.89062486e+06&1.86704234e+06\\ 
\hline
\end{tabular}

\caption{Table 2: Numerical data for the correlation function
$\left\langle T\bar{T}\left( x\right) T\bar{T}\left( 0\right) \right\rangle$.}%
\label{Tab2}%
\end{table}%


\begin{thebibliography}{12}


\bibitem{DN} G. Delfino, G. Niccoli,
hep-th/0407142, Nucl. Phys. B707:381 (2005). 


\bibitem{AZ_VEVTT} A.B. Zamolodchikov, hep-th/0401146, (2004).


\bibitem{Z-Int} A.B. Zamolodchikov,  Adv. Stud. in Pure Math. 19, 641
(1989).

\bibitem{BPZ} A.A. Belavin, A.M.Polyakov, A.B. Zamolodchikov,  Nucl. 
Phys. B 241, 333 (1984).

\bibitem{Z-LY} Al.B. Zamolodchikov,  Nucl. Phys. B348, 619, (1990).

\bibitem{LZ} S. Lukyanov, A. Zamolodchikov, Nucl. Phys. B493, 571
hep-th/9611238, (1997).

\bibitem{FFLZZ} V. Fateev et al., Nucl. Phys. B540, 587, hep-th/9807236,
(1999).

\bibitem{FLZZ} V. Fateev et al., Nucl. Phys. B516, 652, hep-th/9709034, 
(1998).

\bibitem{Smirnov} F. Smirnov, Formfactors in completely integrable models of
quantum field theory, World Scientific, (1992).

\bibitem{Z-TBA} Al.B. Zamolodchikov, Nucl. Phys.
B342, 695, (1990).

\bibitem{Z-RSG} Al.B. Zamolodchikov, Int. J. Mod. Phys., A10, 1125, (1995).

\bibitem{BBLPZ} A.A. Belavin et al.,  
hep-th/0309137, Nucl. Phys. B676, (2004).


\end{thebibliography}
\end{document}